  \documentclass{aa}
   \usepackage{graphicx}
\begin{document}

  % \voffset 3 true cm

   \title{Chromospherically young, kinematically old stars}

   \author{H. J. Rocha-Pinto\inst{1,3} 
          \and B. V. Castilho\inst{2} 
          \and W. J. Maciel\inst{1}}

   \offprints{H. J. Rocha-Pinto, \email{helio@virginia.edu}}

   \institute{Instituto Astron\^omico e Geof\'{\i}sico (USP), 
              Av. Miguel Stefano 4200, 04301-904 S\~ao Paulo SP, 
              Brazil\\  
              \email{maciel@iagusp.usp.br} 
            \and
              Laborat\'orio Nacional de Astrof\'{\i}sica, C.P. 21, 37500-000 
              Itajub\'a MG, 
              Brazil\\ \email{bruno@lna.br}
            \and
              Depart. of Astronomy, University of Virginia, P.O. Box 3818, Charlottesville, VA 22903-0818, USA\\
            \email{helio@virginia.edu}
}
   
   \titlerunning{Chromospherically young, kinematically old stars}
   \authorrunning{H.J. Rocha-Pinto et al.}

   \date{Received date; accepted date}
   
   \abstract{
   We have investigated a group of stars known to have low chromospheric ages, 
   but high kinematical ages. Isochrone, chemical and lithium ages are estimated for 
   them. The majority of stars in this group show lithium abundances much 
   smaller than expected for their chromospheric ages, which is interpreted 
   as an indication of their old age. Radial velocity measurements in the literature 
   also show that they are not close binaries. The results suggest that they can be 
   formed from the coalescence of short-period binaries. Coalescence rates, calculated 
   taking into account several observational data and a maximum theoretical time scale 
   for contact, in a short-period pair, predict a number of coalesced stars similar to what 
   we have found in the solar neighbourhood.
      \keywords{stars: late-type -- stars: chromospheres -- Galaxy: evolution --
                solar neighbourhood}
   }

   \maketitle

\section{Introduction}

The chromospheric activity of a late-type star is frequently interpreted as 
a sign of its youth. Young dwarfs show high rotation 
rates, and the interaction between rotation and outer envelope convection is expected to drive the chromospheric 
activity. 
Nevertheless, not only young single stars present high rotation rates. Close and contact binaries can keep 
high rotation over several billion years. In such stars, the rotational angular momentum loss is balanced 
by the proximity of the stars in the system, which results in transfer of
orbital angular momentum to the rotational spins.  
Thus in chromospheric activity surveys aimed at late type stars, we expect to find two classes of objects: 
young stars and chromospherically active binaries. Sometimes, a star suspected of being young can be instead 
a spectroscopic binary (Soderblom et al. {\cite{soder98}), not yet investigated by radial velocity surveys. 

The chromospheric activity surveys by the Mount Wilson group are directed at, but not only at, late type stars.
The two surveys that comprise the bulk of a sample used by some of us in the derivation of chemodynamical 
constraints on the evolution of the Galaxy (Rocha-Pinto et al. \cite{paperI}, \cite{paperII}) were based on solar-type stars, 
in a spectral range from F8 V to K4 V. Chromospherically active binaries were generally avoided, since the 
surveys investigate the chromospheric activity in single stars. Due to this, the division of these surveys into 
two classes, of active and inactive stars, corresponds closely to an age segregation. 

This was very well demonstrated by Soderblom (\cite{soder90}, see also Jeffries \& Jewell \cite{JJ}), who studied the 
kinematics of active and inactive stars. 
The active stars are concentrated in a region of low velocities in a space velocity diagram, as expected 
for young objects; on the other hand, the inactive stars are scattered in this diagram, just like old stellar 
populations. Few active stars do not follow this rule, showing considerably high velocities. Soderblom calls 
attention to them, but interpret them as possible runaway stars.
Rocha-Pinto et al. (\cite{paperIII}) have increased the number of active stars with spatial velocities to 145. Several 
of these stars, show velocities which are inconsistent with their presumed age.  

The term CYKOS (acronym for chromospherically young, kinematically old stars) is applied here to all 
chromospherically active stars which, in a velocity diagram, present velocity components greater than the  
expected value for such stars, irrespective of the fact that this object is an undiscovered close binary, a 
runaway star or another kind of object. 

This paper presents several newly identified CYKOS, and proposes an explanation for some of them. It is organized 
as follows: in section 2, we present the sample and the method used to define a CYKOS. Section 3 analyses 
what is presently known about their ages from chromospheric and isochrone age measurement methods. In section 
4, we show that these objects have very low Li abundances compared to other stars with the same temperature. 
A critical review of the literature about some individual CYKOS follow in section 5. Finally, in section 6, 
we propose that a small number of these objects could probably have been formed by coalescence of binaries.

   \begin{table*}
      \caption[]{Identified CYKOS. The columns list the stellar name, MK type, $(B-V)$, chromospheric index, chromospheric and 
      isochrone ages, heliocentric spatial velocities $U$, $V$ and $W$, and the components, according to which 
      the object was classified as CYKOS. In the sixth cloumn, the remarks `ZAMS', `red MS' and `sd' refer, respectively, to 
        stars near the zero age main sequence, in the red part of the main sequence, 
        or that are subdwarfs according to their position, and for which no age determination was 
        possible.}
         \label{idcroj}
         \begin{flushleft}
    {\halign{%
    \hfil#\hfil & \hfil#\hfil & \quad\hfil$#$\hfil & \quad\hfil$#$\hfil & \quad\hfil$#$\hfil 
    & \quad\hfil$#$\hfil & \quad\hfil$#$\hfil & \quad\hfil$#$\hfil & \quad\hfil$#$\hfil & 
    \quad\hfil$#$\hfil & \quad\hfil$#$\hfil \cr 
    \noalign{\hrule\medskip}
    HD/BD & MK & (B-V) & \log R'_{\rm HK} & {\rm chrom. age} & {\rm isoch. age} & U & V & W & {\rm criteria} \cr
    \noalign{\medskip\hrule\medskip}
  5303  & G3: V+ & 0.71 & -4.03  &   0.38  &  2.30 &  81\pm 5   &   -58\pm 3   &   -13\pm 1   &  {U}{V}\phantom{W} \cr
  7983  & G2 V & 0.59 & -4.75  &   7.91  &  18.0 &  -138\pm 11  &   -72\pm 6   &   -51\pm 5 &  {U}{V}{W} \cr
 13445  & K1 V & 0.77 & -4.74  &   2.34  &  {\rm red MS} &   98\pm 0.7   &   -75\pm 1   &   -25\pm 2   &  {U}{V}{W} \cr
 16176  & F5 V & 0.48 & -4.73  &   2.07  &  2.0 &  25\pm 2   &   -46\pm 2   &   -22\pm 1   &  \phantom{U}{V}\phantom{W} \cr
 20766  & G2.5 V & 0.64 & -4.65  &   2.09  &  {\rm red MS} &  71\pm 0.5   &   -47\pm 1   &    16\pm 1   &  {U}{V}\phantom{W} \cr
 39917  & G8 V & 0.76 & -4.05  &   0.28  &  1.3 & -61\pm 6   &   -31\pm 2   &   -19\pm 2   &  {U}\phantom{V}\phantom{W} \cr
 51754  & G0 & 0.57 & -4.56  &   3.35  &  &  -194\pm 17  &   -139\pm 13  &    3\pm 1    &  {U}{V}\phantom{W} \cr
 65721  & G6 V & 0.74 & -4.67  &   2.86  &  {\rm red MS} & -37\pm 2   &   -39\pm 2   &    28\pm 1   &  \phantom{U}{V}\phantom{W} \cr
 74385  & K1 V & 0.91 & -4.55  &   0.93  &  {\rm red MS} &  15\pm 0.4   &   -21\pm 2   &   -30\pm 0.5   &  \phantom{U}\phantom{V}{W} \cr
 88742  & G1 V & 0.62 & -4.69  &   2.53  &  10.0 & 36\pm 0.5   &   -45\pm 1   &  -4\pm 1  &  \phantom{U}{V}\phantom{W} \cr
 89995  & F6 V & 0.46 & -4.74  &   4.70  &  2.6 &  55\pm 2   &   -42\pm 2   &   -17\pm 3   &  \phantom{U}{V}\phantom{W} \cr
 103431 & dG7 & 0.76 & -4.68  &   1.74  &  {\rm red MS} &  68\pm 4   &   -39\pm 2   &   -13\pm 2   &  {U}{V}\phantom{W} \cr
 106516 & F5 & 0.46 & -4.65  &   6.43  &  7.4 &  -54\pm 1   &   -74\pm 2   &   -59\pm 2   &  \phantom{U}{V}{W} \cr
 120237 & G3 IV-V & 0.58 & -4.75  &   4.16  &  13.8 &  47\pm 2   &   -57\pm 2   &    -3\pm 1   &  \phantom{U}{V}\phantom{W} \cr
 123651 & G0/G1 V & 0.53 & -4.74  &   5.39  &  13.0 &  29\pm 4   &   -10\pm 4   &   -33\pm 2   &  \phantom{U}\phantom{V}{W} \cr
 131582 & K3 V & 0.96 & -4.73  &   2.79  &  {\rm red MS} &  66\pm 2   &   -67\pm 2   &    15\pm 2   &  {U}{V}\phantom{W} \cr
 131977 & K4 V & 1.11 & -4.49  &   0.21  &  {\rm red MS} & -49\pm 2   &   -22\pm 1   &   -32\pm 1   &  \phantom{U}\phantom{V}{W} \cr
 144872 & K3 V & 0.96 & -4.74  &   2.89  &  {\rm red MS} & -70\pm 1   &    1\pm 1    &    -2\pm 2   &  {U}\phantom{V}\phantom{W} \cr
 149661 & K2 V & 0.81 & -4.58  &   0.83  &  {\rm red MS} &  2\pm 2    &    -1\pm 0.4   &   -30\pm 1   &  \phantom{U}\phantom{V}{W} \cr
 152391 & G8 V & 0.76 & -4.39  &   0.28  &  {\rm red MS} & -85\pm 2   &   -112\pm 2  &    8\pm 1    &  {U}{V}\phantom{W} \cr
 165401 & G0 V & 0.63 & -4.65  &   3.92  &  18.9 &  79\pm 2   &   -90\pm 1   &   -40\pm 1   &  {U}{V}{W} \cr
 183216 & G2 V & 0.60 & -4.62  &   0.50  &  1.8 &  40\pm 2   &   -46\pm 1   &    0\pm 1    &  \phantom{U}{V}\phantom{W} \cr
 189931 & G1 V & 0.60 & -4.64  &   0.76  &  {\rm ZAMS} &  41\pm 2   &   -50\pm 1   &    1\pm 1    &  \phantom{U}{V}\phantom{W} \cr
 196850 & G0 & 0.57 & -4.64  &   1.31  &  {\rm red MS} &  0\pm 0.4    &   -22\pm 2   &   -33\pm 1   &  \phantom{U}\phantom{V}{W} \cr
% 204121 & F5 V & 0.42 & -4.66  &   0.95  &  2.0 &  -5\pm 3   &   -21\pm 4   &   -41\pm 3   &  \phantom{U}\phantom{V}{W} \cr
 209100 & K4.5 V & 1.06 & -4.56  &   0.39  &  {\rm red MS} &  80\pm 1   &   -41\pm 0.2   &    4\pm 1    &  {U}{V}\phantom{W} \cr
 219709 & G2/G3 V & 0.65 & -4.62  &   1.14  &  8.5 &  25\pm 1   &   -42\pm 1   &   -10\pm 0.1   &  \phantom{U}{V}\phantom{W} \cr
 230409 & G0 & 0.70 & -4.70  &   8.78  &  {\rm sd} &-133\pm 16   &   -129\pm 13  &   -13\pm 1   &  {U}{V}\phantom{W} \cr
+15 3364 & G0 & 0.63 & -4.43  &   0.49  & 8.7 & -59\pm 3    &   -22\pm 2   &    2\pm 1    &  {U}\phantom{V}\phantom{W} \cr% 
%+42 2163 & K1 V & 0.82 & -4.71  &   3.09  & {\rm sd} &  105   &   -109  &   -76   &  {U}{V}{W} \cr
+51 1696 & sdG0 & 0.55 & -4.42  &   3.51  &   & 228\pm 23   &   -277\pm 33  &    53\pm 1   &  {U}{V}{W} \cr
  \noalign{\medskip\hrule}}} 
         \end{flushleft}
   \end{table*}

\section{Identification of CYKOS}

The CYKOS can be identified by a diagram of spatial velocities ($U\times V$ 
or $W\times V$) showing only active stars, in analogy to their first discovery by Soderblom. 
Not all objects identified in a $U\times V$ diagram are also identified in a $W\times V$ diagram, and vice versa. 
We expect that CYKOS showing high velocities in more than one component are really peculiar objects, and 
not just stars having a component velocity in the tail of the distribution. 

      \begin{figure*}
      \sidecaption
      \includegraphics[width=12cm]{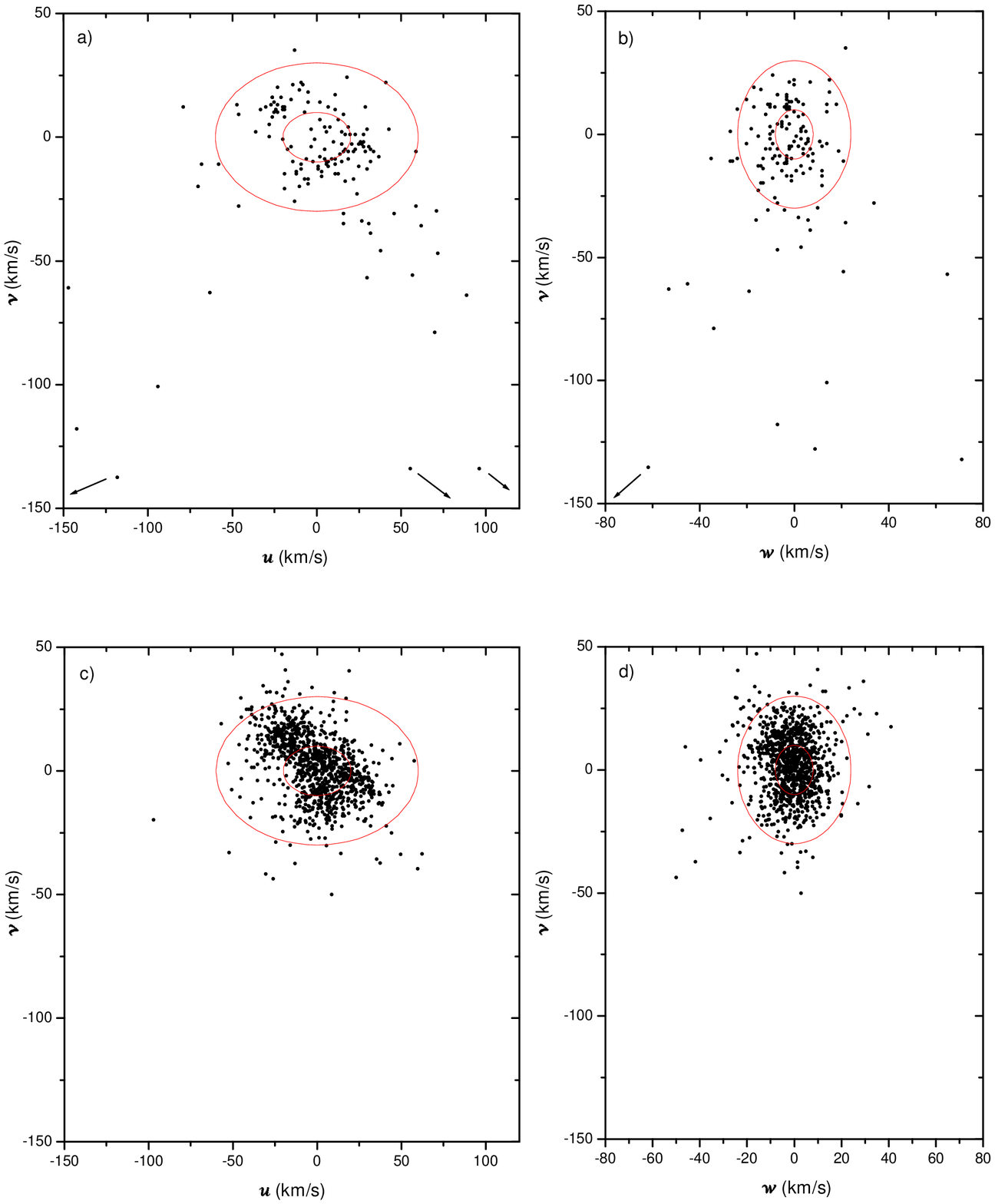}
      \caption{
         Panels a and b: spatial velocity diagrams for 145 active stars. The ellipses have semiaxes equal to 
         $\sigma$ and $3\sigma$, where $\sigma$ is the velocity dispersion for stars with ages lower  
         than 1 Gyr (Rocha-Pinto et al. \cite{paperIII}) in the corresponding velocity components. For these  
         dispersions, we use $\sigma_u= 20$ km/s, $\sigma_v=10$ km/s and $\sigma_w=8$ km/s. Objects 
         located beyond the outer ellipses, corresponding to $3\sigma$, are considered CYKOS. Panels c and d: 
         spatial velocity diagrams for 1023 A dwarfs, taken from the sample of Palou\v s (\cite{palous}). In all 
         panels, we have used spatial velocities corrected for the solar motion (Mihalas \& Binney \cite{mihalas}).
          }
      \label{uvwvcroj}
      \end{figure*}

Some CYKOS also appear clearly in an age--velocity diagram, so that, in principle, this diagram  
could also be used to identify them. However, the uncertainty 
in the chromospheric ages can make some inactive stars appear young 
in such a plot, and our main purpose is not the study of young inactive stars, but rather of active stars that 
are kinematically old.

We have used as our primary source the sample containing 145 active stars built by Rocha-Pinto et al. 
(\cite{paperIII}). This sample is composed by all active
stars from 
 Rocha-Pinto et al. (\cite{paperII}) for which radial velocity
measurements are available in the literature. 
The heliocentric velocities are calculated with the equations provided by 
Johnson \& Soderblom (\cite{johnsod}). 
 
Figs.\ref{uvwvcroj}a and b show the $U\times V$ and $W\times V$ diagrams for these stars. 
The velocities were corrected for the solar motion $(U_\odot, V_\odot, W_\odot) = (-9, 11, 6)$ km s$^{-1}$ 
according to 
Mihalas \& Binney (\cite{mihalas}). 
In the plots, the semiaxes of the inner and outer ellipses are equal to 1$\sigma$ and 3$\sigma$, respectively,  
where $\sigma$ is the velocity dispersion of stars with ages lower than 1 Gyr. We have adopted  
$\sigma_U= 20$ km/s, $\sigma_V=10$ km/s and $\sigma_W=8$ km/s, which correspond to the velocity 
dispersion of the youngest stellar population (ages 0 to 1 Gyr), according to Rocha-Pinto 
et al. (\cite{paperIII}), and have considered as CYKOS all objects located beyond the outer ellipses.
The velocity dispersions for the youngest population are in good agreement with other determinations in the 
literature. The values for $\sigma_U$ range from 11.7 km/s (Meusinger et al. \cite{meusinger}) to 23 km/s 
(Cayrel de Strobel \cite{cayrel}). The agreement is closer for $\sigma_V$, ranging from 5.9 km/s 
(Meusinger et al. \cite{meusinger}) to 10.1 km/s (Cayrel de Strobel \cite{cayrel}), and 
$\sigma_W$, ranging from 8 km/s (Wielen \cite{wielen}) to 9.7 km/s (Cayrel de
Strobel \cite{cayrel}).

In Tab. (\ref{idcroj}), we list the thirty stars identified as CYKOS. 
The first column gives the name of the 
star, followed by the spectral type, $(B-V)$ colour, chromospheric activity index, 
chromospheric and isochrone ages, as commented 
upon later, and the $U$, $V$ 
and $W$ heliocentric velocities, which are given with formal uncertainties. The last column shows the 
velocity criteria used in order to classify the object as a CYKOS.

The results also show that CYKOS are generally more apparent in $V$ than in the other velocity components, 
as expected for a kinematically old stellar population: the $V$ component is systematically 
more negative than in the case of normal active stars, which are 
presumably young. This is an effect of the asymmetrical drift. The 
stars are likely to acquire increasing random velocities with respect to the local standard of rest, 
due to subsequent encounters with giant molecular clouds. 
In $U$ and $W$, there will be a symmetric increase in the velocity dispersion 
and we would not expect the CYKOS to be much different from the normal stars, from the 
consideration of these velocity components only.

The peculiar character of these objects can be seen from Figs. \ref{uvwvcroj}c and d, where the
same diagrams are shown for 1023 A dwarfs, taken from the large compilation by Palou\v s (\cite{palous}). 
Since A dwarfs are very young stars, due to their maximum life expectancy, their kinematical properties 
are consistent with those of very young late-type dwarfs. We do find stars having velocity components greater 
than 3$\sigma$, where we have used the same velocity dispersions used in Figs. \ref{uvwvcroj}a and b. 
Nevertheless, these A dwarf outliers have smaller velocities than the average velocities of the CYKOS 
as shown in Figs. \ref{uvwvcroj}a and b. Aroung 4\% 
of the A dwarfs are outliers, while this number goes to 20\% in the case of the late-type dwarfs. 
For the A stars, these outliers probably represent the tail of the velocity distribution, which is reinforced by the fact that 
their $V$ distribution is nearly symmetrical, contrary to what happens for the CYKOS.

\section{What age measurement methods tell us}

The chromospheric ages of half of the objects listed in Tab. (\ref{idcroj}) are lower or 
similar to 2 Gyr.
This can be seen from 
the fourth and fifth columns of the Table, which give the chromospheric activity index 
$\log R'_{\rm HK}$ and the chromospheric ages in Gyr, respectively. The chromospheric index was defined by 
Noyes et al. (\cite{annoyes}) and the calculation of chromospheric ages was discussed by 
Rocha-Pinto \& Maciel (\cite{RPM98}). 

There are 9 stars in Table \ref{idcroj} whose chromospheric age is larger than 3 Gyr. Strictly speaking, 
they cannot be considered as `chromospherically young', and are listed in view of their activity levels 
($\log R'_{\rm HK}>-4.75$), 
which traditionally indicate a young age (Soderblom \cite{soder90}). Five of them 
have $\log R'_{\rm HK} \le -4.70$. Given that the error in $\log R'_{\rm HK}$ is expected to be around 0.04 dex, 
it is possible that these stars are inactive, rather than active stars. On the other hand, at least one of them presents significant 
X-ray emission (\object{HD 89995}). Also, we must take into account that a high chromospheric age could be caused 
by the metallicity of the star. 
Metal-poor CYKOS are also chromospherically older, due to the metallicity dependence of the chromospheric 
age (Rocha-Pinto \& Maciel \cite{RPM98}). In fact, some of these 9 stars are metal-poor, in comparison with the majority 
of the other stars.  However, since the metallicity introduces an additional source of error 
in the calculation of the chromospheric age, it is not unlikely that these objects could be younger than 
what is shown in Table \ref{idcroj}. For instance, we have remarked that even some of these older CYKOS 
have velocities considerably larger than 
the mean velocity of their coeval stars (this is particularly true for the CYKOS having chromospheric age between 
2 and 4 Gyr).  For these reasons only, we  
have decided to keep them in the sample. 

The basic problem deserving an explanation is why the chromospheric 
ages of these objects are low (sometimes very low), while their kinematic ages are high (sometimes very high). 
What can be said about their ages from other methods?

%     \begin{table}[h]
%      \caption[]{Isochrone ages of CYKOS. The remarks `ZAMS', `red MS' and `sd' refer respectively to 
%        stars that have fallen in the zero age main sequence, in the red part of the main sequence, 
%        or that are 
%        subdwarfs according to their position, and for which no age determination was 
%        possible.}
%         \label{HRcroj}
%         \begin{flushleft}
%    {\halign{%
%    \hfil#\hfil & \quad\hfil#\hfil & 
%    \qquad\qquad\hfil# \hfil & \quad\hfil#\hfil \cr
%    \noalign{\hrule\medskip}
%    HD/BD & age (Gyr) & HD/BD & age (Gyr) \cr
%    \noalign{\medskip\hrule\medskip}
%5303 & 2.3 & 131977 & red MS\cr
%7983  & 18.0 &  144872 & red MS\cr 
%13445 & red MS & 149661 & red MS  \cr
%16176 & 2.0 & 152391 & red MS\cr
%20766 & red MS & 165401 & 18.9  \cr
%39917 & 1.3 & 183216 & 1.8\cr
%65721 & red MS & 189931 & ZAMS \cr
%74385 & red MS & 196850 & red MS \cr
%88742 & 10.0 & 204121 & 2.0  \cr
%89995 & 2.6 & 209100 & red MS\cr
%103431 & red MS & 219709 & 8.5  \cr
%106516 & 7.4 & 230409 & sd\cr
%120237 & 13.8 & +15 3364 & 8.7 \cr
%123651 & 13.0 & +42 2163 & sd  \cr
%131582 & red MS   & \cr
%  \noalign{\medskip\hrule}}} 
%         \end{flushleft}
%   \end{table}

The objects in Tab.(\ref{idcroj}) have a broad metallicity distribution. Most of these stars have 
[Fe/H] between $-0.40$ and $+0.20$, but 11\% of them have photometric metallicities lower than $-0.60$ dex. 
From the point of view of chemical evolution, they have a broad age range, with averages around 3-5 Gyr if we 
adopt the age--metalliticy relation given by Rocha-Pinto et al. (\cite{paperII}).

For the calculation of isochrone ages, we have used the isochrones by  
Bertelli et al. (\cite{bertelli}). The age of the star was calculated by interpolation in each isochrone. 
A final interpolation, taking into account the ages at several metallicities (that is, several isochrone grids), 
uses the real stellar metallicity to find the stellar age. The results are shown in the sixth column of 
Tab. \ref{idcroj}. 

These ages are somewhat rough and some care must be taken when interpreting these results.  
The reason for this is that the $m_1$ deficiency, 
present in active stars (see Rocha-Pinto \& Maciel \cite{RPM98}, and references therein), 
hinders the determination of accurate stellar parameters by photometric indices. In our case,  
$\log T_{\rm eff}$, which is estimated from the equations given by Olsen (\cite{olsen84}), can 
be miscalculated. Three stars do not have parallaxes measured by HIPPARCOS, 
or have {\it uvby} colours outside the range covered by the calibrations by 
Olsen, and do not have isochrone ages in Table \ref{idcroj}. 

Nearly 40\% of the stars lie in the red part of the main sequence, for which no age determination is possible. 
The remaining stars are distributed nearly equally between young (7 stars having less or about 2 Gyr)
and old stars (8 stars with ages greater than 7 Gyr). There are no preferred ages for these stars.

Our results show that, in spite of some of CYKOS having low isochrone ages, 
others can be very old. The cooler stars in the red MS can have 
very different ages, since they are in a colour range where no perceptible evolution in the HR diagram is 
visible.

\section{Lithium in CYKOS}

     \begin{table*}
      \centering
      \caption[]{Observations of CYKOS and active stars.}
         \label{journal}
         \begin{flushleft}
    {\halign{%
    \hfil#\hfil & \quad\hfil#& \hskip 3pt \hfil#\ &
    \hskip 3pt \hfil# & \qquad\hfil#&
    \hskip 3pt \hfil#& \hskip 3pt \hfil#\hfil& \quad\hfil#\hfil &
    \quad\hfil#\hfil  &\quad\hfil#\hfil &\quad\hfil#\hfil
    &\quad\hfil#\hfil&\quad\hfil$#$\hfil&\quad\hfil#\hfil&\quad\hfil#\hfil\cr
    \noalign{\hrule\medskip}
    HD/BD & \multispan3 {\underbar{R.A. (2000)}} & \multispan3
    {\underbar{Dec. (2000)}} &
    J.D. & exp. & T$_{\rm eff}$ & $\log g$ & $\xi_T $ & {\rm [Fe/H]} & N(Li) & v$_r$\hfil \cr
          & h & m & s & $\degr$ & $'$ & $''$ &  (2451000 +) & (s) & (K) & & (km/s) & & & (km/s)\cr
    \noalign{\medskip\hrule\medskip}
 870 & 00 & 12 & 51 & $-$57 & 54 & 48 & 037.782 & 900 & 5350 & 4.40 & 0.5 & -0.12 & 0.50 & \hfil $-$2\cr
1237   & 00 & 16 & 04 & $-$79 & 51 & 02 & 384.785 & 660 & 5250 & 4.33 & 0.7 & 0.00 & 1.90 & $-$3\cr% 
%3329   & 00 & 35 & 32 & $-$00 & 30 & 17 & 384.795 & 500 & 6490 & 3.97 & 1.8 & -0.20 & 2.20 & $-$14\cr
13445  & 02 & 10 & 15 & $-$50 & 50 & 00 & 037.836 & 900 & 5270 & 4.50 & 0.8 & -0.10 & 0.50 & \hfil +53\cr
17051  & 02 & 42 & 31 & $-$50 & 48 & 12 & 384.837 & 450 & 6040 & 4.25 & 1.0 & 0.10 & 2.40 & +4\cr
%17084  & 02 & 43 & 25 & $-$37 & 55 & 40 & 384.814 & 1300 & & +93\cr
20766  & 03 & 17 & 36 & $-$62 & 35 & 04 & 037.850 & 600 & 5715 & 4.30 & 1.0 & -0.20 & 0.50 & \hfil +9\cr
22049  & 03 & 32 & 59 & $-$09 & 27 & 31 & 384.832 & 220 & 5215 & 4.83 & 1.3 & -0.15 & 0.75 & \cr
106516 & 12 & 15 & 10 & $-$10 & 17 & 54 & 384.398 & 600 & 6190 & 4.20 & 1.0 & -0.55 & $< 1.20$ & +8\cr
%115383 & 13 & 16 & 47 &    09 & 25 & 17 & 384.407 & 500 & \cr
124580 & 14 & 15 & 38 & $-$44 & 59 & 55 & 037.410 & 1200 & 5845 & 4.28 & 1.5 & -0.20 & 2.75 & \hfil +4~\cr
131977 & 14 & 54 & 32 & $-$21 & 11 & 28 & 033.452 & 900 & 4585 & 4.58 & 1.0 & 0.20 & $< -1.00$ & \hfil +26\cr
138268 & 14 & 15 & 58 & $-$44 & 59 & 55 & 033.502 & 2400 & 5975 & 4.40 & 1.0 & 0.30 & 2.90 & \hfil $-$59\cr
  "    &    &    &    &       &    &    & 037.426 & 1200 &      &      &     &      & 2.90 &  \cr
149661 & 16 & 36 & 19 & $-$02 & 19 & 13 & 037.442 & 600 & 5235 & 4.50 & 1.0 & 0.05 & $< -1.00$ & \hfil $-$14\cr
152391 & 16 & 53 & 01 & $-$00 & 00 & 22 & 037.455 & 900 & 5450 & 4.37 & 1.2 & 0.00 & 1.10 & \hfil +44\cr
154417 & 17 & 05 & 16 &    00 & 42 & 25 & 037.481 & 666 & 5970 & 4.35 & 1.3 & -0.10 & 2.80 & \hfil $-$18\cr
165401 & 18 & 05 & 37 &    04 & 39 & 42 & 037.468 & 900 & 5755 & 4.21 & 1.0 & -0.40 & 0.50 & \hfil $-$120\cr
174429 & 18 & 53 & 05 & $-$50 & 10 & 49 & 037.528 & 1320 & 5100 & 4.10 &    &       & 3.00 & \hfil $-$10\cr
181321 & 19 & 21 & 29 & $-$34 & 58 & 56 & 037.567 & 780 & 5975 & 4.30 & 1.5 & 0.10 & 3.10 & \hfil $-$19\cr
185124 & 19 & 37 & 46 & $-$04 & 38 & 48 & 037.579 & 480 & 6760 & 4.20 &     &      & 2.80 & \hfil $-$33\cr
189931 & 20 & 04 & 02 & $-$37 & 52 & 15 & 037.587 & 900 & 5865 & 4.35 & 1.0 & 0.10 & 2.15 & \hfil $-$44\cr
202628 & 21 & 18 & 25 & $-$43 & 20 & 05 & 384.697 & 660 & 5750 & 4.24 & 1.0 & 0.05 & 2.15 & \hfil +9\cr
202917 & 21 & 20 & 49 & $-$53 & 01 & 58 & 384.672 & 1800 & 5555 & 4.25 & 1.6 & 0.05 & 3.35 & $-$1\cr
%204121 & 21 & 26 & 27 &    01 & 06 & 20 & 384.655 & 540 & 6530 & 4.22 & 1.5 & 0.00 & 2.40 & \hfil +1\cr
206667 & 21 & 44 & 44 & $-$42 & 07 & 46 & 384.709 & 1000 & 5950 & 4.22 & 1.0 & 0.00 & 2.40 & +14\cr
209100 & 22 & 03 & 21 & $-$56 & 47 & 09 & 384.664 & 400 & 4660 & 4.90 & 1.5 & 0.10 & 0.15 & \hfil $-$44\cr
217343 & 23 & 00 & 18 & $-$26 & 09 & 05 & 384.723 & 900 & 5755 & 4.44 & 1.3 & 0.00 & 3.20 & $-$9\cr
221231 & 23 & 31 & 00 & $-$69 & 04 & 29 & 384.751 & 800 & 5910 & 4.42 & 1.3 & 0.00 & 2.95 & +2\cr
%222335 & 23 & 39 & 50 & $-$32 & 44 & 20 & 384.773 & 740 & & $-$24\cr
223913 & 23 & 53 & 40 & $-$65 & 56 & 55 & 384.762 & 660 & 5985 & 4.50 & 1.0 & 0.15 & 2.65 & +18\cr
+15 3364 & 18 & 07 & 18 &  15 & 56 & 54 & 037.491 & 1800 & 5685 & 4.20 & 0.8 & 0.00 & 0.00 & \hfil +24\cr
   "   &    &    &    &       &    &    & 037.504 & 1800 &      &      &     &      & 0.00 & \cr
  \noalign{\medskip\hrule}}} 
         \end{flushleft}
   \end{table*}

Although Li depletion and production in stars are processes not
completely
well understood, in some cases the Li abundance could be used as a youth
indicator.
If CYKOS are young objects, as suggested by their chromospheric
activity, they
must present high Li abundances. On the other hand, if they show very
depleted Li,
they must be evolved objects, irrespective of what their magnetic
activity might tell us.

\subsection{Observations}

We have obtained spectra for 28 stars, including CYKOS and normal active
stars (used
as reference objects). The observations were carried out at the
Laborat\'orio
Nacional de Astrof\'{\i}sica (LNA, Brazil) in two observing runs, in
August 1998
and July 1999. The spectra were obtained with the coud\'e spectrograph at
the 1.6m
telescope, using a SITe CCD of 1024$\times$1024 pixels with
24$\mu$m$\times$24$\mu$m
pixel size, and a grating of 1800~l/mm yielding a resolution over two
pixels of
$\approx$ 24,000 covering the wavelength range $\lambda\lambda$ 6640--6780
\AA.

The data were reduced using standard tasks of IRAF package. Spectra from
different
exposures were added by weighting them with (S/N)$^2$. The final S/N
ratio for the
stars are in the range of 100 to 200. The spectrum of a hot star obtained with the
same
configuration was inspected for telluric lines. The log of observations
is reported
in Table \ref{journal} together with some other information. The
radial velocities
listed were calculated using a set of unblended atomic lines in the same
data.  The error in the radial velocity is 2 km/s.

\subsection{Spectra of normal stars and CYKOS} 

      \begin{figure*}
      \centering
      \includegraphics[height=21cm]{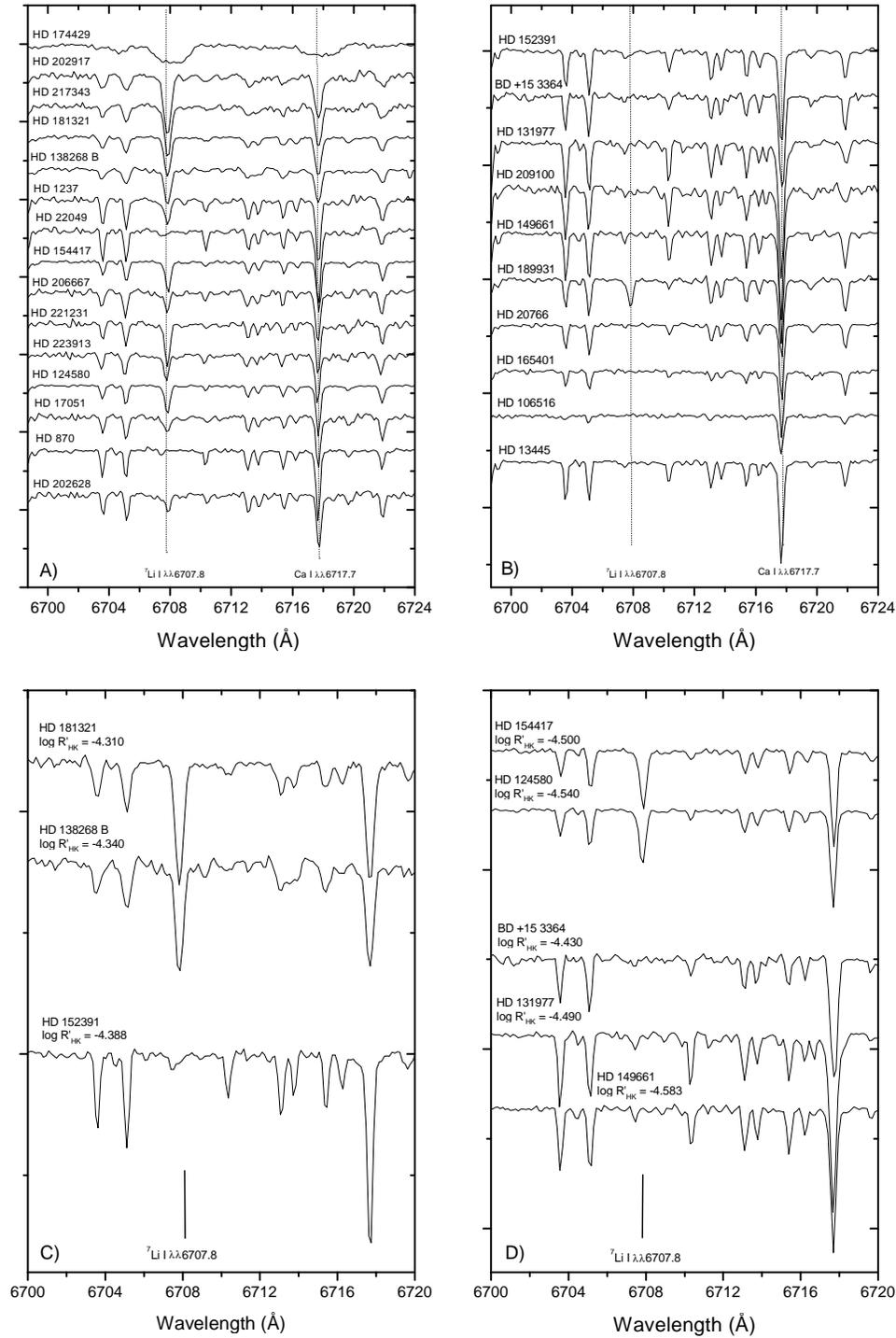}
      \caption{
Spectra for active stars in the Li region ($\lambda$ 6707 \AA). 
Panel {\it a} presents normal young stars, while panel {\it b} 
presents CYKOS. The spectra are ordered according to the chromospheric activity, 
with the most active stars in the upper part of the figure. With few exceptions 
it is seen that the Li line is present in all normal active stars, and its equivalent 
width becomes smaller towards less active stars. 
This behaviour is totally absent from the spectra of CYKOS. In panels {\it c} and {\it d} 
we compare spectra of stars having the same activity levels and, supposedly, the same age. 
The upper spectra are for normal stars, in which there is Li. In the bottom spectra, 
of CYKOS, the Li line is absent.
}
      \label{specs}
      \end{figure*}

In Fig.~\ref{specs}a, we show spectra in the Li region ($\lambda$ 6707 \AA) 
for some objects taken as normal. 
The spectra in this plot are ordered according to the chromospheric activity, the lowest spectra being 
that of the least active star. 

The chromospheric activity order must represent approximately an age order: the most active 
objects are supposed to be younger than the least active ones. In fact, by inspecting the plot 
we can see a gradual increase in the equivalent width of the Li line, in going up the Figure, from \object{HD 202628} 
to \object{HD 174429}. The spectrum of \object{HD 185124} was too broadened to allow some comparison with 
other stars, and was not included in the figure. 
The presence of lithium and a high chromospheric activity are classical 
youth indicators in late-type stars. The exceptions are few, and do not contradict this idea. 
\object{HD 22049} ($\equiv \epsilon$ Eri) is a BY Dra variable, and possibly older   
than its chromospheric activity suggests. \object{HD 870} has  
an activity level near that of the Vaughan-Preston gap, and could be considered as an inactive star, 
observed during a maximum of activity. 
% \object{HD 3229} is the most intriguing case. 
%Having $\log R'_{\rm HK} = -4.58$, we expected to find an equivalent width for the lithium line 
%similar to that of \object{HD 124580}. It must not be an old object: Duncan et al. (\cite{duncan}) present  
%statistics about 450 observations for this star from 1966 to 1982, and its activity level remains relatively constant. 
%The chromospheric age calculated for this star is 2.43 Gyr, using the equations by Rocha-Pinto \& Maciel 
%(\cite{RPM98}). Although it is not so young as some stars of the plot, it should present some Li. 

The spectra of CYKOS are presented in Fig.~\ref{specs}b. 
When we examine the same chromospheric activity sequence amongst the CYKOS, we find nothing similar to 
that found in normal stars. The Li line is only present in the spectra of \object{HD 189931}. 
In the others there is no trace of Li. The comparison can be done more properly in the bottom panels of 
this figure, where we show normal stars and CYKOS with the same chromospheric activity levels, and supposedly, 
the same age.

\subsection{Stellar Parameters and Models}

Stellar parameters (T$_{\rm eff}$, $\log g$, [Fe/H]) for the program stars were
first derived from the {\it ubvy} photometry and the classical relation
$\log g_*=  4.44 + 4\log T_*/T_{\odot} + 0.4(M_{\rm bol}-4.74) + \log
M_*/M_{\odot}$,
using the same procedure described in Castilho et al. (\cite{cast00}). 

The metallicities of the program stars were then redetermined using
curves of
growth of Fe I lines where the updated code RENOIR by M. Spite was
employed. The stellar parameters used, with the spectroscopic metallicity 
that we have found, are listed in Table \ref{journal}. For 
\object{HD 185124} and \object{HD 174429} a metallicity determination was 
not possible, due to the line broadening.  The error in [Fe/H] is 0.10 dex.

Model atmospheres employed have been interpolated in tables computed
with
the MARCS code by Edvardsson et al. (\cite{edv93}).

\subsection{Li abundances} \label{Analysis}

The Li depletion must not be identical in all observed stars, since it
depends
on factors such as mass and metallicity. Even the equivalent
width of
the $^7$Li resonance doublet ($\lambda$ 6707.8\AA) depends strongly on
the temperature
of the star (Castilho et al. \cite{cast00}). Therefore, a strong Li line does not
always
warrant a high lithium abundance.  The stars in Fig.~\ref{specs} do not have the same 
temperature, and those plots shall be taken as illustrative comparisons. More accurate estimates of the Li age
must be
done by considering Li abundances rather than the equivalent widths.

Spectrum synthesis calculations were used to fit the observed spectra
of the stars listed in Tab. \ref{journal}. 
The calculations of synthetic spectra were carried out using a revised
version of the code described in Barbuy (1982), where molecular lines of
C$_2$ (A$^3 \Pi$-X$^3 \Pi$), CN red (A$^2 \Pi$-X$^2 \Sigma$) and TiO
$\gamma$
(A$^3 \Phi$-X$^3 \Delta$) systems are taken into account.
The oscillator strengths adopted are the laboratory values obtained by
Fuhr et al. (\cite{fuhr}), Martin et al. (\cite{martin}), Wiese et al. (\cite{wiese}). When
these were not available, we have used those given by Spite et al. (\cite{spite}) or Barbuy
et al. (\cite{barbuy1999}) obtained by inverse solar analysis. Solar abundances are
adopted from Grevesse \& Sauval (\cite{gs99}). 
The derived lithium abundances $N({\rm Li})$ of the observed stars are given in 
column 10 of Tab. \ref{journal}.

      \begin{figure*}
     \centering
      \includegraphics[height=22cm]{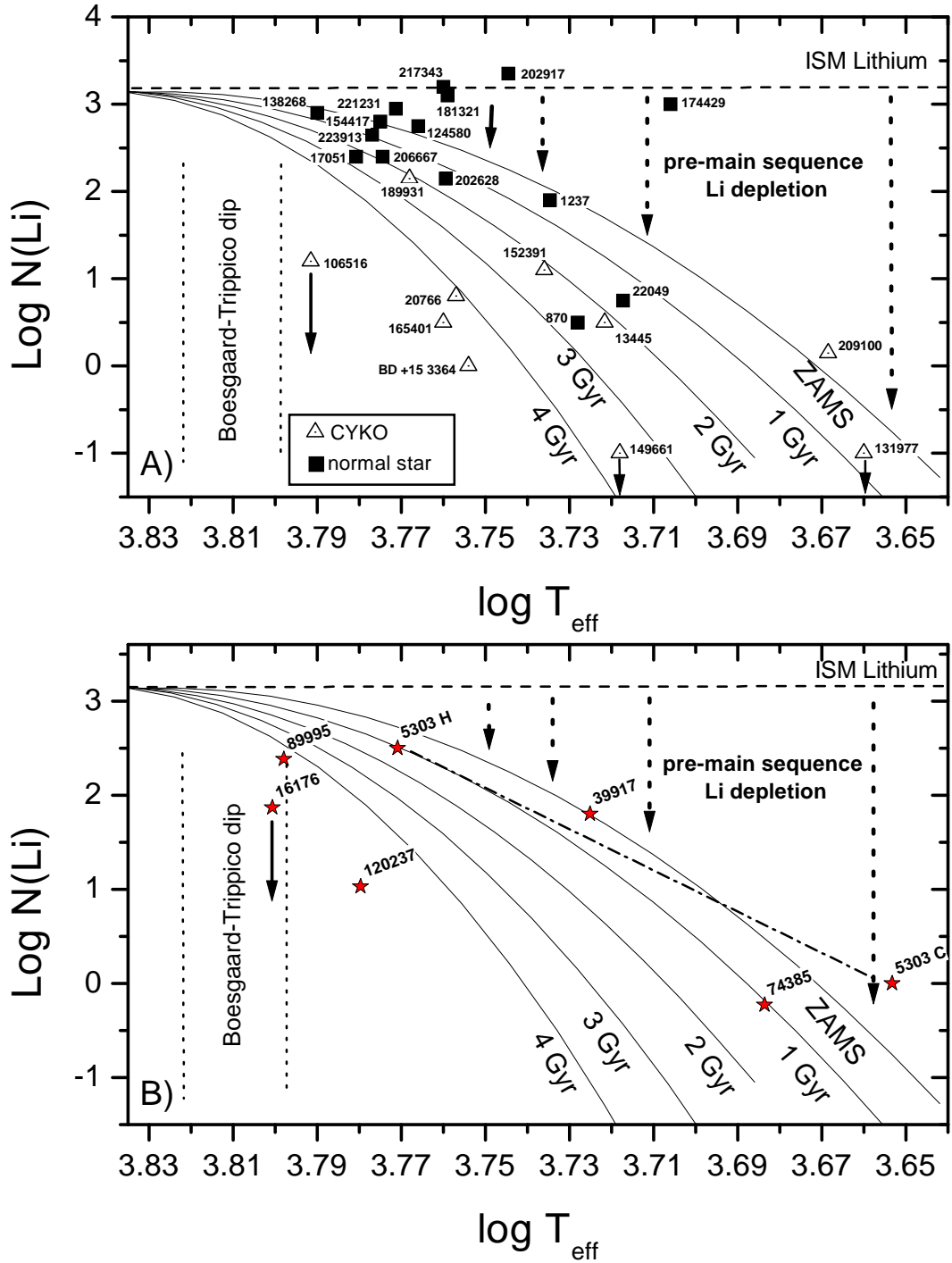}
      \caption[]{
Li depletion diagram for late-type stars, according to Soderblom (\cite{soder83}). 
Panel {\it a}: stars that we have observed; 
panel {\it b}: CYKOS with Li abundances from the literature. In this figure, 
the dashed horizontal line at the 
top indicates the Li abundance in the interstellar medium. Vertical arrows 
correspond to the Li depletion before the 
main sequence. Several curves indicate the Li depletion expected for 
stars of a given age. The vertical dotted lines in the 
left part of the plots correspond to the region of the Boesgaard--Trippico dip.
}
      \label{littest}
      \end{figure*}

For the determination of the lithium ages of these stars, we have considered 
the lithium depletion diagram of Soderblom (\cite{soder83}). 
Fig.~\ref{littest} presents a Li depletion diagram as a function of the stellar effective temperature, according 
to Soderblom (\cite{soder83}). The diagram is an approximation for the calculation of Li ages. The dashed horizontal 
line 
corresponds to the Li abundance in the interstellar medium, and the curves indicate the expected depletion 
as a function of the stellar age. Vertical arrows show the expected lithium depletion before the ZAMS. The vertical 
dotted lines mark the region corresponding to the Boesgaard--Trippico dip 
(cf., Boesgaard \& Trippico \cite{boes}), where depletion is not linked with age, but probably with 
the internal structure of the star. 
The filled squares in the top panel of Fig. \ref{littest} are the normal young stars of Table \ref{journal}, 
while the triangles show the CYKOS. The lower panel of Fig. \ref{littest} shows the Li depletion 
diagram of a few CYKOS (stars) whose abundances are from the literature (section 5).  
It is clear that normal young stars have lithium abundances very similar to 
that of the interstellar medium. That is, in these stars there was no depletion. The CYKOS, 
on the other hand, have Li ages greater than 2 Gyr. 

\object{HD 870} is the only exception amongst normal young stars. As mentioned before, this could be an inactive star, 
included in the sample by an error in $\log R'_{\rm HK}$ (for instance, it has been observed in 
the H an K Ca II lines only once, by Henry et al. \cite{HSDB}). \object{HD 189931}, in spite of having N(Li) $\ga 2.0$, 
has the same Li age as \object{HD 152391} and \object{HD 13445}.

We conclude that CYKOS present Li ages higher than the ages estimated by their chromospheric activity, 
although in 
better agreement with their kinematic properties.

\section{Data on individual objects}

The initial suggestion by Soderblom (\cite{soder90}) that the CYKOS can be runaway stars seems very 
unrealistic. If these 
stars were young, they should present Li abundances typical of 
their youth. Their high kinematic age is a clear 
indication of their older status. An initial hypothesis that can be tested 
is that they could be chromospherically 
active binaries, which can be tested by searching for radial velocity variations. 
Before giving a general explanation, 
it is important to see what is known about the stars in Tab.~\ref{idcroj}. 

\par\smallskip
\par\noindent \object{HD 5303} $\equiv$ CF Tuc
\par\smallskip
This is a G0 V + K4 IV RS CVn-type spectroscopic binary (Strassmeier et al. \cite{strass}), 
with an orbital period of 2.80 days. Li abundances were measured by Randich et al. (\cite{randich94}) 
for both components. In Fig.~\ref{littest}b, both components are linked by a dot-dashed line. Note that  
this line presents roughly the same slope of the curves with the same Li age.
\par\smallskip
\par\noindent HD 13445 $\equiv$ HR 637
\par\smallskip
The index $\log R'_{\rm HK}=-4.74$ indicates that this star could be an inactive star. A comparison between 
 H$\alpha$ fluxes (Pasquini \& Pallavicini \cite{pasquini}) with other stars is ambiguous:  
HR 13445 presents fluxes similar to HD 42807 and HD 81997 (active stars) and to HD 4308 and the Sun 
(inactive stars). Favata et al. (\cite{favata}) have measured N(Li) $<-0.24$, one order of magnitude lower 
than our measured value. Queloz et al. (\cite{queloz}) have found  
a planet with 4 Jupiter masses at a distance of 0.11 AU from the star. 
\par\smallskip
\par\noindent HD 16176 $\equiv$ HR 756
\par\smallskip
This is classified as F5. Balachandran (\cite{bal}) measured N(Li) $<1.87$. The depletion would be high for this 
spectral type, if due to age alone. However, the star is found within the Boesgaard--Trippico dip, and this 
low lithium abundance cannot be considered as indicative of old age.
\par\smallskip
\par\noindent HD 20766 $\equiv \zeta_1\,\,{\rm Ret}\equiv$ HR 756
\par\smallskip
This is a G2.5 V star, visual companion of HD 20807, with an angular separation of 307$''$. Wooley (\cite{wooley}) suggests   
that they are members of the $\zeta$ Her moving group, which has an isochrone age of a few 
billion years. da Silva \& Foy (\cite{licio}) 
measured a metallicity typical of population I stars and, therefore, criticized the hypothesis by Johnson et al. 
(\cite{johnson}) that the pair is composed of subdwarfs. There is no sign of radial velocity variability in either stars (da 
Silva \& Foy \cite{licio}).
\par\smallskip
\par\noindent HD 39917 $\equiv$ SZ Pic
\par\smallskip
This is a RS CVn chromospherically active star (Strassmeier et al. \cite{strass}), having an orbital period of 
4.80 days. However, Mason et al. (\cite{mason}) have not detected a companion within  
$\Delta V \le 3.0$ and angular separation between $0.035''$ and $1.08''$. The  
lithium abundance was measured by Randich et al. (\cite{randich93}), and is consistent with the expected abundance at 
the ZAMS (see Fig. \ref{littest}). 
\par\smallskip
\par\noindent HD 65721
\par\smallskip
G6 V variable star, ROSAT source (H\"unsch et al. \cite{huensch98}). Mason et al. (\cite{mason}) have not detected 
a companion within  
$\Delta V \le 3.0$ and angular separation between $0.035''$ and $1.08''$. 
\par\smallskip
\par\noindent HD 74385 
\par\smallskip
Dwarf K1 V. Favata et al. (\cite{favata}) have measured N(Li) $<-0.23$.
\par\smallskip
\par\noindent HD 88742 $\equiv$ HR 4013
\par\smallskip
G1 V star, ROSAT source (H\"unsch et al. \cite{huensch99}). Mason et al. (\cite{mason}) have not detected a 
companion within  
$\Delta V \le 3.0$ and angular separation between $0.035''$ and $1.08''$.
\par\smallskip
\par\noindent HD 89995 $\equiv$ HR 4079
\par\smallskip
F6 V star, ROSAT source (H\"unsch et al. \cite{huensch99}). The lithium abundance measured by Balachandran 
(\cite{bal}; N(Li) $ = 2.38$) 
is high compared to some stars, but small for the temperature of this star (6280 K). 
The star is located within the Boesgaard--Trippico dip, where depletion is uncorrelated with age.
\par\smallskip
\par\noindent HD 103431 
\par\smallskip
It is a dG7 star, visual companion of HD 103432, angular separation of 73.2$''$. Constant radial 
velocity during a time span of 2499 days (Duquennoy \& Mayor \cite{duc91}).
\par\smallskip
\par\noindent HD 106516 $\equiv$ HR 4657
\par\smallskip
F5 V spectroscopic binary with period of 853.2 days (Latham et al. \cite{latham}). 
Lithium abundance was measured by Lambert et al. (\cite{lambert}), N(Li) $<1.32$.
The star is also located within the Boesgaard--Trippico dip. 
Edvardsson et al. (\cite{edv93}) calculate an isochrone age 
of 5.37 Gyr. Fuhrmann \& Bernkopf (\cite{fuhrmann}) suggest that this star 
is a field blue straggler, having a chemical composition 
and kinematics typical of thick disk stars, 
in spite of having an age typical of thin disk stars. 
\par\smallskip
\par\noindent HD 120237 $\equiv$ HR 5189
\par\smallskip
It is classified as G3 IV-V. There is no indication of a companion within  
$\Delta V \le 3.0$ and angular separation between $0.035''$ and $1.08''$ (Mason et al. \cite{mason}). 
The lithium abundance was calculated by Randich et al. (\cite{randich99}), N(Li) $=1.03$. 
The depletion seems substantial for the temperature of this star.
\par\smallskip
\par\noindent HD 123651
\par\smallskip
There is no indication of a companion within  
$\Delta V \le 3.0$ and angular separation between $0.035''$ and $1.08''$ (Mason et al. \cite{mason}). 
\par\smallskip
\par\noindent HD 131977
\par\smallskip
K4 V visual binary, with angular separation of 20$''$. The companion is HD 131976. Its x-ray emission level 
is moderate (Wood et al. \cite{wood}), but there seems to be no 
doubt about its activity (Robinson, Cram \& Giampapa \cite{robin}). 
Duquennoy \& Mayor (\cite{duc88}) have found no invisible companion for this star.
\par\smallskip
\par\noindent HD 149661 $\equiv$ 12 Oph $\equiv$ V2133 Oph $\equiv$ HR 6171
\par\smallskip
K2 V variable of BY Dra type (Petit \cite{petit}). It was detected by ROSAT in EUVE with moderate intensity  
(Tsikoudi \& Kellett \cite{tsil}). Habing et al. (\cite{habing}) report a Vega-like protoplanetary disk, 
but the presence of cirrus 
during the observation has somewhat made this finding inconclusive. Young, Mielbrecht \& Abt (\cite{young}) 
and Tokovinin (\cite{tokovin}) 
have found a constant radial velocity for this star, and McAlister et al. 
(\cite{mcallister}) did not detect the presence of an unseen 
companion by using speckle interferometry.
\par\smallskip
\par\noindent HD 152391 $\equiv$ V2292 Oph
\par\smallskip
G8 V variable star of BY Dra type (Petit \cite{petit}). Detected by ROSAT in EUVE with moderate intensity 
(Tsikoudi \& Kellett \cite{tsil}). 
Constant radial velocity during a timespan of 3387 days (Duquennoy \& Mayor \cite{duc91}).
\par\smallskip
\par\noindent HD 165401
\par\smallskip
It is a G0 V, relativaly metal-poor star ([Fe/H] $\sim -0.50$ dex). 
The index $\log R'_{\rm HK}$ that we have used refers to a sole observation (Duncan et al. \cite{duncan}). 
We have considered the possibility that this star is inactive, but 
its emission in H$\alpha$ (Herbig \cite{herbig}) seems  
consistent with its $\log R'_{\rm HK}$. There is no unseen companion with a magnitude diference lower than 2.5 mag 
and angular separation greater than 3 AU, according to speckle interferometry (Lu et al. \cite{lu-patinadora}). 
Abt \& Levy (\cite{abtlevy}) have found a radial velocity variability of $\pm 5$ km/s, but recent investigations do not 
confirm this result (Duquennoy \& Mayor \cite{duc91}; Abt \& Willmarth \cite{abtwill}). Curiously, radial velocities measured for 
this star during the sixties and seventies yield values homogeneously around $v_r\approx -114$ km/s, while all recent 
studies yield values around $v_r\approx -120$ km/s. 
\par\smallskip
\par\noindent HD 196850 
\par\smallskip
G2 V star, with constant radial velocity during a time span of 3995 days (Duquennoy \& Mayor \cite{duc91}).
\par\smallskip
%\par\noindent HD 204121 $\equiv$ HR 8205
%\par\smallskip
%F5 V ROSAT source (H\"unsch et al. \cite{huensch98}). The lithium abundance measured by Balachandran 
%(\cite{bal}; N(Li) $ = 2.39$) 
%is high compared to some stars, but lower for its temperature (6537 K). 
%The star is also located within the Boesgaard--Trippico dip.
%\par\smallskip
\par\noindent BD +15 3364
\par\smallskip
Solar-metallicity G0 V star. Duquennoy \& Mayor (\cite{duc91}) have found a constant radial velocity during a time span 
of 2264 days. Carney (\cite{carney}) found no photometric variability characteristic of an unseen companion. 
\medskip

\section{The nature of CYKOS}

\subsection{Coalescence of close binaries}

The majority of objects considered in the previous section are undoubtedly active, have lithium abundances 
lower than that of stars with 2 Gyr of age, and some show no indication of radial velocity variations.  
They are thus single objects, most probably old. 

As we have seen, some of them are chromospherically active binaries. Their chromospheric activity  
results from the synchronization of the orbital with the rotational motion. This is the reason why they have 
a low chromospheric age, but a higher age from the point of view of their kinematics. Nevertheless, in the case of 
single stars, there is no known mechanism that could store angular momentum to be used later 
by the star. The anomalously low lithium abundance, together with the high velocity components, are 
hardly interpreted as a consequence of anything but an old age. Even the BY Dra variables in the sample must be 
single stars, and not chromospherically active binaries, since there is no indication of variability in $v_r$ for 
them. Amongst BY Dra stars are found binaries as well as single stars (Eker \cite{eker}). 
Soderblom (\cite{soder90}) 
has estimated a kinematic age of 1-2 Gyr for them, contrary to the expectation that these stars could keep  
high activity levels at advanced ages. The low kinematic age must be understood as an evidence that the BY Dra-type 
variability is not exclusive of chromospherically active stars, but could be generated by the intensity of the magnetic 
activity itself amongst low-mass stars, binaries or not. 

Poveda et al. (\cite{allen},b) also have found several CYKOS amongst UV Ceti stars, which are known as young low-mass stars.
The authors suggest that these objects could be red stragglers, a low-mass analogous of the well-known blue stragglers. 
According to this hypothesis, the red stragglers would be produced by the coalescence of two low-mass stars (with about 0.5 $M_\odot$ 
each) originally in a short-period binary pair. According to them, these objects would be {\it stragglers} in a velocity diagram, 
compared to other stars. However, being more rapid, they would not be stragglers in the sense they are in the 
HR diagram. Thus, the name `field blue stragglers' includes not only the idea 
about their origin, but also their present location out of a cluster. 
Note that the same denomination was used by Fuhrmann \& Bernkopf (\cite{fuhrmann}).

The coalescence scenario was already considered as a classic explanation for the formation of 
blue stragglers. Two works (van't Veer \& Maceroni \cite{vant}; St\c epie\'n \cite{stepien}) that investigate  
the coalescence of short-period binaries into a single star predict the formation of low-mass blue stragglers. 
St\c epie\'n (\cite{stepien}) has even shown that the coalescence is more easily attained for low-mass binaries 
(each having around 0.6 $M_\odot$, forming a star with 1.2 $M_\odot$) than for more massive stars that 
originate the 
classic blue stragglers in open clusters. The formation of a low-mass blue straggler could be achieved 
within 2.5 Gyr, for systems having an initial orbital period of 2 days. 

The properties of a supposed low-mass field blue straggler would be tightly similar to that of some CYKOS, as we will 
see in what follows. 

In short-period binaries, we expect the occurrence of synchronization between the orbital and rotational periods. 
For low-mass stars, the magnetic activity is strong, and increases the angular momentum loss.  
When both orbital and rotational periods are synchronized, the rotational angular momentum loss 
occurs at the expenses of the orbital angular momentum. 
As a result, the period decreases, the components rotate more rapidly, and become closer to each other, eventually 
becoming contact binaries, as those of W UMa-type. 

Rasio \& Shapiro (\cite{rasio}) simulate systems like these, using the technique of smooth particle hydrodynamics. The authors 
show that once the contact is achieved, these systems are dynamically unstable and can rapidly coalesce 
into a single object, having a high rotation rate. The events related to this coalescence can produce 
an extense outer envelope that would make the star appear like a pre-main sequence star. According to the 
authors, the coalescence can occur in a time scale of a few hours, once the dynamical instability is set. The envelope 
is kept gravitationally bound to the star, which eventually contracts towards thermal equilibrium. Mass loss, in this 
event, would be minimal.

In a coalescence of two low-mass stars (0.5 $M_\odot$ each), the resulting star must present a mass similar to that 
of the Sun and a high rotation rate. This rotation rate, together 
with the convection in the outer stellar atmosphere, would 
produce a copious chromospheric activity, similar to the one found in very young stars. In the case of low-mass stars that 
have not ignited hydrogen in their cores, the just formed single star would be similar in many respects to a young star,
positioning in the zero-age main sequence, like the blue stragglers. However, this star would inherit the same 
velocity components of the binary pair from which it was formed. Thus, due to the time before the coalescence, the 
velocity components are not similar to that of a young star. We would have a star almost in everything young, but 
kinematically old. 
The lithium abundance is also one of the few tracks that can show the real nature of these objects. 
In spite of not burning hydrogen considerably, stars having around 0.5 $M_\odot$ 
are highly convective, and Li burning is very efficient in them. Blue Stragglers like these should present small or no 
Li abundance (Pritchet \& Glaspey \cite{pritchet}; Glaspey et al. \cite{glaspey}), since they would be formed by older objects.

A criticism that could be made is that, if CYKOS are field blue stragglers, formed during the coalescence of 
low-mass short-period binaries, why do their spectra not present very broad lines, as it is expected in stars 
with high rotation rates? 
The same problem occurs for blue stragglers in open and globular clusters, that do not rotate more rapidly 
than young normal stars. Leonard \& Livio (\cite{leonard}) suggest that the majority of the angular momentum is stored 
in the extense disk that is formed around just formed blue stragglers. The central object would expand, due to the 
thermal energy resulting from the coalescence, and contracts toward thermal equilibrium, in a time scale lower than 
10$^7$ years, similar to that of pre-main sequence stars.
Thus, we do not find CYKOS with broader lines simply because we do not expect that they all have been formed 
within the last 0.5 Gyr. 

Seven amongst our stars seem to fit well within these criteria: HD 20766, HD 106516, 
HD 131977, HD 149661, HD 152391, HD 165401, BD +15 3364. Other stars are suspect, but the 
information is insufficient 
to characterize them as low-mass blue stragglers. One of the stars (HD 189931) seems 
to be a real young single object with high velocity 
components (see also Tab. \ref{idcroj}). Nevertheless, there is little
information 
 in the literature that could test this hypothesis.
The other stars need more data to investigate whether they are field blue stragglers, 
chromospherically active stars or runaway stars. 

\subsection{Coalescence Rate}

The formation rate of these objects can be calculated from considerations about the initial mass function, star formation 
rate, initial period distribution in binaries and the time scale for contact. The time needed for the coalescence, once contact is 
achieved, is very small compared to the time scale for contact (Rasio \& Shapiro \cite{rasio}), and we will consider it negligible.

      \begin{figure*}[ht]
      \centering
      \includegraphics[width=12cm]{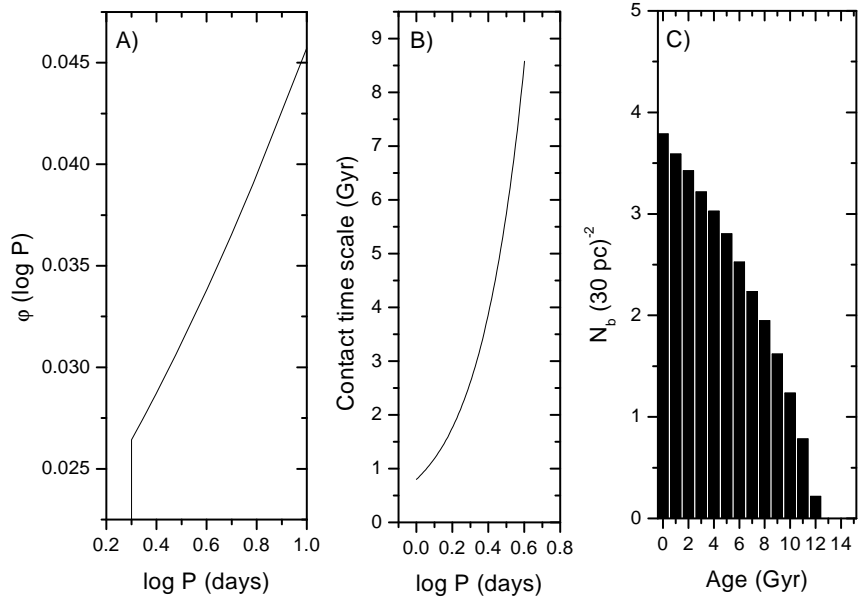}
      \caption[]{
         Calculations for the formation rate of population I field blue stragglers having masses between 
         0.8-1.2 $M_\odot$. Panel a: initial period distribution, adapted from  
         Duquennoy \& Mayor (\cite{duc91}); panel b: time scale for contact; panel c: number of 
         field blue stragglers, with mass in the considered range, formed within a sphere of 30 pc radius 
         centred at the Sun.  }
      \label{bluestra}
      \end{figure*}

It can be shown that the total number of stars, having masses between $m_1$ and $m_2$, that becomes contact binaries in 
$t$ is 
\begin{equation}
   N_b(t) = b\int^{{m_2}}_{{m_1}}\int^\infty_{0.3}\int^{0.5}_0  {{\cal P}}(\mu,\log P, m, t) 
   \,{\rm d}\mu\,{\rm d}\log P\,{\rm d}m
\label{Nbintot}
\end{equation}
\begin{equation}
{{\cal P}}(\mu,\log P, m, t) = f(\mu)\varphi(\log P)\psi(t-\tau_{\mu P m})\phi(m),
\label{NcalP}
\end{equation} 
where $\tau_{\mu P m}$ is the time scale for contact for a binary, having initial period 
in the range $(\log P, \log$ $P 
+\Delta\log P)$, secondary-to-total mass ratio between $(\mu,\mu+\Delta\mu)$ and total mass between $(m,m+\Delta m)$,
$\varphi(\log P)$ is the initial period distribution, $f(\mu)$ is the distribution of mass ratio, $\psi(t)$ is the 
star formation rate, and $\phi(m)$ is the initial mass function. 

For the computation of the equation above, we will consider $\varphi(\log P)\approx 0.018+0.027\log P$, with a cutoff  
for $\log P < 0.3$, which approximates fairly well the initial period distribution of Duquennoy \& Mayor (\cite{duc91}), for the region 
$0.3<\log P<1.0$, where $P$ is in days. We have assumed $f(\mu)=24\mu^2$ (Matteucci \& Greggio \cite{greggio}), 
$m_1=0.8 M_\odot$ and $m_2=1.2 M_\odot$, which correspond to pairs having total masses equal to the mass range for G 
dwarfs.

The function $\tau_{\mu P m}$ is very complicated. St\c epie\'n (\cite{stepien}) published calculations for 
only three binary configurations: $1 M_\odot+1 M_\odot$, $1 M_\odot+0.65M_\odot$ and 
$0.6 M_\odot+0.6 M_\odot$. The last of these has total mass equal to $m_2$, but calculations for lower 
total masses are not published, neither for different $\mu$. However, St\c epie\'n 
says that configurations for lower 
total masses achieve contact in a time scale lower than that for the system $0.6 M_\odot+0.6 M_\odot$. 
We consider that $\tau_{\mu P m}$ can be approximated by a function 
$\tau_P$, which depends only on the initial period, and that is given by the time for contact for the 
system  $0.6 M_\odot+0.6 M_\odot$. Note that being $\tau_P$ a maximum time scale, our estimates will be 
a little underestimated, since the total number of systems that have achieved contact $\tau_P$ after the formation of the 
binary will be greater that the calculated number, due to the number of binaries with lower total masses that 
achieve contact more rapidly. Also, due to the initial mass function, the number of systems with lower 
total masses must be higher. Taken these into account, Eq. \ref{Nbintot} reduces to 
$N_b(t)\ga cF(t)$, where 
\begin{equation}
c=b\int^{1.2}_{0.8}\phi(m)\,{\rm d}m
\label{c}
\end{equation}
and
\begin{equation}
F(t)=\int^\infty_{0.3}\varphi(\log P)\psi(t-\tau_P)\,{\rm d}\log P,
\label{ft}
\end{equation}
since the integral in $\mu$ is unity.

For the calculation of $\tau_P$, we have considered Fig. 2 of St\c epie\'n (\cite{stepien}). The time scales for  
contact were fitted by $\tau_P=0.8e^{4\log P}$. We have assumed a constant star formation rate, in spite of 
the evidences for its non-constancy (Rocha-Pinto et al. \cite{paperII}), since we are only interested in the 
magnitude of the coalescence rate.

In Fig. \ref{bluestra}a, the initial period distribution is shown. The time scale for contact is shown in panel b. The 
number of field blue stragglers already formed was calculated for a sphere of radius 30 pc around the Sun, which 
correspond approximately to the volume within which our sample is nearly complete. 
This rate is shown in Fig. \ref{bluestra}c.  
Integrating the data in this panel, we have a total number of 28 blue stragglers already formed 
in this sphere. 

The calculations are very rough, as can be seen from the approximations considered. However, 
the number of objects formed in the sphere is similar in order of magnitude to the number of 
field blue stragglers in our sample (including suspect objects). This reinforces our hypothesis for the 
nature of these objects.

From these results, we believe to have found population I field blue stragglers. Poveda et al. (\cite{allen},b) 
arise a hypothesis that cannot be tested with the same probability level as ours, since they do not 
have an indication of old age for their stars, besides their kinematics. 

The traditional method used for finding a blue straggler, based on its position in the HR diagram, is 
not possible to 
be used for a population I field blue straggler. The search for objects with
strong chromospheric activity, or little Li, or with high velocity, also do
not allow their identification, since normal stars present these properties.
Only by the intersection of several properties we have found objects which 
apparently can be best explained by this scenario.

It is worth mentioning an independent research on ultra-lithium-deficient halo stars 
recently published by Ryan et al. (\cite{ryan}), which conclude that 
such stars can have the same origin of the blue stragglers, being their lower mass the 
only significant difference between them.                       

\begin{acknowledgements}
      The authors are indebted to an anonymous referee, who has made important suggestions to an 
      earlier version of this paper. 
      The SIMBAD database, operated at CDS, 
      Strasbourg, France, was used throughout this research. We acknowledge support by FAPESP and CNPq. 
\end{acknowledgements}

\end{document}